\documentstyle[pre,aps,preprint]{revtex}

\input{epsf}

\begin{document}

\draft

\title{Influence of weak anchoring on flow instabilities in nematic 
liquid crystals}

\author{O.~S. Tarasov$^{\footnotesize{1,2}}$, A.~P
Krekhov$^{\footnotesize{1,2}}$ and L. Kramer$^{\footnotesize{1}}$}

\address{$^{1}$Institute of Physics, University of Bayreuth, D-95440 Bayreuth,
Germany}

\address{$^{2}$Institute of Molecule and Crystal Physics, Russian Academy of
Sciences, 450025 Ufa, Russia}

\date{\today}

\maketitle

\begin{abstract}
We analyse the homogeneous instabilities in a nematic liquid
crystal subjected to plane steady Couette or Poiseuille flow in the case
when the director is pre-aligned perpendicular to the
flow plane taking into account weak anchoring at the confining
surfaces.
The critical shear rate decreases for decreasing anchoring 
strength and goes
to zero in the limit of torque-free boundary conditions.
For Poiseuille flow one has two types of instability
depending on the values of the azimuthal ($W_a$) and polar ($W_p$)
surface anchoring strengths.
The critical line in $(W_a,W_p)$ space which separates the two
instabilities regimes is obtained.
\end{abstract}



\section{Introduction}

During the last decade the study of surface
anchoring in nematic liquid crystals (NLCs)
for different types of confining substrates has attracted much attention.  
In particular, a change of anchoring
strength strongly influences the orientational behaviour and dynamic response
of NLCs under external electric and magnetic fields. This changes the switching
times, which play an
important role for applications (see, e.g., \cite{chigrinov}). The hydrodynamic
flow  is a crucial ingredient for the dynamic response and switching
characteristics of liquid crystal devices.
  
The anchoring characteristics can also be studied in
orientational phenomena induced by hydrodynamic flow. In fact, recently the
surface orientational transition caused by oscillatory shear
flow was found \cite{khazkre} and the  influence of weak anchoring on the
linear response of the NLC to oscillatory flows was studied
\cite{naskre}.   Up to now the study of
orientational bulk instabilities in NLCs under hydrodynamic flow have been
restricted to the case of strong surface anchoring (fixed director orientation
at the confining plates) \cite{lesadv,edvpmbook,obzor}.

In this paper the influence of surface anchoring on the homogeneous
instabilities in NLCs subjected to {\em steady flow} of Couette and
Poiseuille type is studied theoretically in the case when the 
director at the bounding plates is oriented {\em perpendicular to the
flow plane}. This is the simplest geometry because the initial
state  with the director oriented everywhere perpendicular to the flow plane
is,  by symmetry, a solution of the nematodynamic equations for any shear rate.
The state can change only via a symmetry-breaking instability. The type of
instability strongly depends on the sign of the Leslie viscosity coefficient
$\alpha_3$.  The case of negative $\alpha_3$ ({\em flow-aligning} materials)
and strong anchoring of the director was
investigated theoretically by Leslie \cite{lesJPD76} and Dubois-Violette
and Manneville \cite{edvpmbook,mdv,mdv1,mann}.
They have shown that in the absence of external fields the first
instability is a homogeneous one in both, steady
Couette \cite{lesJPD76,mdv} and Poiseuille \cite{mdv1} flows . 
The theoretical results were
found to be in good agreement with experiments of Pieranski and Guyon
and coworkers \cite{pg1,sum1,dgjpm}. 
In Couette flow sufficiently strong additional  magnetic field applied parallel
to the initial director orientation was found to change the type of instability
into a spatially  periodic one where rolls develop \cite{pg1,dgjpm},
in contrast to the case of Poiseuille flow, 
where a magnetic field does not change the type of instability \cite{jpg}. 
Well above the threshold of the homogeneous instability in Poiseuille flow 
rolls were observed to develop in a secondary instability \cite{pg}. 
In the case of positive $\alpha_3$ ({\em non-flow-aligning} materials),
according to the mechanism of Pieranski and Guyon, one has no
homogeneous instability and only rolls are expected \cite{pg1}.
Since for these materials there is also a tumbling motion 
\cite{sum2}, the orientational behaviour of NLCs can be 
quite complex.

Here we focus on {\it flow-aligning} nematics.
Starting from the standard set of nematodynamic equations in the 
Leslie-Ericksen formulation \cite{lesadv} the basic equations 
describing the homogeneous instabilities are presented
(sec.~\ref{conteqs}) taking into account arbitrary surface 
anchoring strength.
Full semi-analytical solutions together with approximate expressions
for the critical shear rate at which the homogeneous instabilities
develop are obtained for steady Couette (sec.~\ref{cf}) and
Poiseuille (sec.~\ref{pf}) flow.
In sec.~\ref{disc} we discuss the results.

%
\section{Basic equations}
\label{conteqs}
We consider a NLC layer of thickness $d$ sandwiched between two parallel
infinite plates. The origin of a Cartesian coordinate system is placed at the 
centre of the layer with the $z$ axis perpendicular to the bounding plates.
Steady Couette flow is generated by one plate (at $z=d/2$) moving with  
constant velocity $V^0$ along the $x$ direction and the other plate (at
$z=-d/2$)  fixed.
Steady Poiseuille flow is obtained when a constant pressure gradient 
$\Delta P/\Delta x$ is applied along the $x$ axis.
The confining plates provide a director orientation along the
$y$ axis. The basic state is
given by the stationary homogeneous solution of the
standard set of nematodynamic equations (Leslie-Ericksen equations
\cite{lesadv})
\begin{eqnarray}
\label{initstate}
{\bf n}^0=(0,1,0) \; , \quad {\bf v}^0=(v_x^0,0,0) \; ,
\end{eqnarray}
where
\begin{eqnarray}
\begin{array}{rclcl}
v_x^0&=&V^0(1/2+z/d) &-&\mbox{for Couette flow }, \\
v_x^0&=&-\frac{\Delta P}{\Delta x}\frac{d^2}{2 \eta_3}(1/4-z^2/d^2)
&-&\mbox{for Poiseuille flow, }
\end{array}
\end{eqnarray}
and $\eta_3=\alpha_4/2$ with $\alpha_4$ the corresponding Leslie viscosity.

In order to investigate the stability of solution (\ref{initstate}) we
linearise the nematodynamic equations \cite{lesadv} 
with respect to perturbations that are homogeneous in the
plane of the layer
\begin{eqnarray}
{\bf n}={\bf n}^0+(n_x,n_y,n_z) \; , \quad 
{\bf v}={\bf v}^0+(v_x,v_y,v_z) \; ,
\end{eqnarray}
where $n_i, v_i$  ($i=x,y,z$) are functions of $z$. We are looking for the
existence of a stationary solution of the linearised nematodynamic equations,
which signals the onset of a stationary (i.e. nonoscillatory) instability. 
The linearised equations are 
\begin{eqnarray}
\begin{array}{lcc}
\eta_1 v_{y_{,zz}}+(\eta_1-\eta_3) (S n_x)_{,z} &=& 0 \; , \\
\alpha_2 S n_z - K_{22} n_{x_{,zz}} &=& 0 \; , \\
\alpha_3 S n_x - K_{11} n_{z_{,zz}} + \alpha_3 v_{y_{,z}} &=& 0 \; ,
\end{array}
\label{baseqn}
\end{eqnarray}
where $S=v_{x_{,z}}^0$ is the shear rate,  
$\eta_1=(\alpha_3+\alpha_4+\alpha_6)/2$ and $K_{11}$, $K_{22}$ are the
elasticity coefficients for ``splay'' and ``twist'' deformations, respectively.
The notation $f_{,z} \equiv df/dz$ is used throughout.

The boundary conditions for the $y$-component of the velocity 
perturbation are 
\begin{equation}
\label{bcvy}
v_y(z=\pm d/2) = 0 \; .
\end{equation}
The surface anchoring of the director is described in terms of a surface energy
per unit area $F_s$ which has a minimum when the director at the surface is
oriented along the easy axis (parallel to the $y$ axis in our case).
It is convenient to characterise the surface anchoring by a ``polar''
anchoring strength $W_p$ pertaining to out-of-substrate-plane director
deviations and an ``azimuthal'' anchoring strength $W_a$ related to 
distortions within the substrate plane.
A phenomenological expression for the surface energy $F_s$ can be written 
in terms of an expansion with respect to (${\bf n}-{\bf n}^0$). For small
director deviations from the easy axis one  obtains for the surface energy 
\begin{eqnarray}
F_s = \frac12 W_a n_x^2 + \frac12 W_p n_z^2 \;, \quad W_a>0, W_p>0 \; .
\end{eqnarray}
The boundary conditions for the director perturbations can be
obtained from the surface torques balance equation 
\begin{eqnarray}
\label{bcnxz}
\pm K_{22} n_{x_{,z}} + \frac{\partial F_s}{\partial n_x} = 0 \; ,
\quad
\pm K_{11} n_{z_{,z}} + \frac{\partial F_s}{\partial n_z} = 0 \; ,
\quad
\mbox{for $z=\pm \; d/2$.} 
\end{eqnarray}
Introducing the dimensionless quantities
\begin{eqnarray}
\begin{array}{lcl}
{\tilde z} = \frac{z}{d} \; , \quad 
{\tilde S} = \beta \tau_d S \; , \quad
\tau_d = \frac{(-\alpha_2) d^2}{K_{22}}  \\
V_y=\frac{\beta^2 b \tau_d}{d} v_y \; , \quad
N_x=\beta n_x, \quad N_z=n_z
\end{array}
\end{eqnarray}
with 
\begin{eqnarray}
\beta^2=\frac{\alpha_3}{\alpha_2}\frac{K_{22}}{K_{11}} \frac{1}{b}, \quad
b=\frac{\eta_1}{\eta_3}, 
\end{eqnarray}
Eqs.(\ref{baseqn}) can be rewritten in the form
\begin{eqnarray}
\label{eqnvynxnz}
\begin{array}{lcc}
V_{y_{,zz}} - (1-b) ({\tilde S} N_x)_{,z} &=& 0 \; , \\
{\tilde S} N_z + N_{x_{,zz}} &=& 0 \; , \\
b {\tilde S} N_x + N_{z_{,zz}} + V_{y_{,z}} &=& 0 \; .
\end{array}
\end{eqnarray}
For the shear rate ${\tilde S}$ one has
\begin{eqnarray}
{\tilde S} =& a^2 \; , \; a^2 = \frac{V^0 \tau_d}{d}\beta
\quad &\mbox{ --- for Couette flow }, \label{bascouette} \\
{\tilde S} =& - a^2 z \; , \; 
a^2 = -\frac{\Delta P}{\Delta x}\frac{\tau_d d}{\eta_3}\beta
\quad &\mbox{ --- for Poiseuille flow } \label{baspois}.
\end{eqnarray}
The boundary conditions (\ref{bcvy}), (\ref{bcnxz}) reduce to
\begin{eqnarray}
\label{bcvynxnz}
\pm N_{x_{,z}} + w_a N_x = 0 \; , \quad  
\pm N_{z_{,z}} + w_p N_z = 0 \; , \quad
V_y = 0 \; , &
\mbox{ at } z=\pm \; 1/2, 
\end{eqnarray}
where 
\begin{equation}
w_a=W_a d/K_{22} \; , \quad w_p=W_p d/K_{11} \; .
\label{dimw}
\end{equation}
In the limit of strong anchoring ($w_a$, $w_p \to \infty$) one has
$N_x = N_z = 0$ at $z=\pm1/2$ whereas for torque-free boundary conditions
($w_a$, $w_p \to 0$) $N_{x_{,z}} = N_{z_{,z}} = 0$ at the boundaries.
From (\ref{dimw}) one can see that by changing the thickness $d$,
the dimensionless anchoring strengths $w_a$ and $w_p$ can be varied, the ratio
$w_a/w_p$ remaining fixed.

Solving the system of linear ordinary differential equations 
(\ref{eqnvynxnz}) with 
boundary conditions (\ref{bcvynxnz}) one can obtain the critical 
value of the shear rate $a^2_c$, at which the initial state (\ref{initstate})
loses stability, and determine the influence of anchoring strengths $w_a$
and $w_p$ on $a^2_c$.

%
\section{Couette flow}
\label{cf}
In this case the shear rate ${\tilde S}$ is independent of $z$
(\ref{bascouette}), so that the system (\ref{eqnvynxnz}) can be solved using
standard theory of ordinary differential equations with constant
coefficients. Further, the \{$z \to -z$\} symmetry of Eqs.(\ref{eqnvynxnz}) 
allows for two possible types of solutions:  %
\begin{eqnarray*}
\mbox{I:}&&
\{ V_y \mbox{ - even}; N_x \;, N_z \mbox{ - odd}\} 
\mbox{ - ``odd'' solution }, \\
\mbox{II:}&&
\{ V_y \mbox{ - odd}; N_x \; , N_z \mbox{ - even}\} 
\mbox{ - ``even'' solution }.
\end{eqnarray*}
We will always classify the solutions by the $z$-symmetry of the
$x$-component of the director perturbation $N_x$.
For the {\em odd}  solution one obtains 
\begin{eqnarray}
&&V_y = C_1 (1-b) a \cosh(a z) - C_2 (1-b) a \cos(a z) + C_3 \; ,\nonumber \\ 
&&N_x = C_1 \sinh(a z) + C_2 \sin(a z) \; , \\
&&N_z = -C_1 \sinh(a z) + C_2 \sin(a z) \; . \nonumber
\end{eqnarray}
Taking into account the boundary conditions 
(\ref{bcvynxnz}) the solvability condition for the $C_i$ (''boundary 
determinant`` equated to zero) gives the 
expression for the critical shear rate 
\begin{eqnarray}
\begin{array}{ll}
2 w_a w_p \sinh(a_c/2) \sin(a_c/2) + \\
a_c (w_a+w_p) [ \cosh(a_c/2) \sin(a_c/2) + \cos(a_c/2) \sinh(a_c/2) ] + \\
2 a_c^2 \cosh(a_c/2) \cos(a_c/2) = 0 \; .
\end{array}
\label{odddet}
\end{eqnarray}
For the {\em even} solution one obtains 
\begin{eqnarray}
\label{shevensol}
&&V_y = C_1 (1-b) a \sinh(a z) + C_2 (1-b) a \sin(a z)- C_3 a^2 b z \; ,
\nonumber \\ 
&&N_x = C_1 \cosh(a z) + C_2 \cos(a z) + C_3 \; , \\
&&N_z = -C_1 \cosh(a z) + C_2 \cos(a z) \; . \nonumber
\end{eqnarray}
The boundary  determinant condition now gives 
\begin{eqnarray}
\begin{array}{ll}
a_c^3 b \sinh(a_c/2)\sin(a_c/2) + \\
\frac12 a_c^2 b (w_a+w_p) [ \cosh(a_c/2)\sin(a_c/2) - \cos(a_c/2)\sinh(a_c/2) ] - \\
\{a_c b \cosh(a_c/2)\cos(a_c/2) + \\
  (1-b) 
  [ \sinh(a_c/2)\cos(a_c/2) + \sin(a_c/2)\cosh(a_c/2)]\} w_a w_p = 0 \; .
\end{array}
\label{evendet}
\end{eqnarray}
Note, that the expressions (\ref{odddet}) and (\ref{evendet}) are both 
symmetric under exchange of $w_a$ and $w_p$. This results from the
fact that $N_x$ and $N_z$ have the same symmetry. 
We found that for MBBA material parameters at 25$^o$ \cite{kn} ($b=0.58$)
the critical shear rate  for the {\em odd}  mode is  systematically higher
than for the {\em even} mode, so there is no transition between them under
variation of the surface anchoring. For strong anchoring ($w_a=w_p=\infty$) the
expression for the critical shear rate of the {\em even} solution
(\ref{evendet}) recovers the result obtained by Leslie
\cite{lesJPD76}.
Weak surface anchoring reduces the critical shear rate 
compared to the case of strong boundary conditions.
In the limit of one of the surface anchoring strengths going to zero
($w_a\to 0$ or $w_p\to 0$) one has $a^2_c \to 0$.
In  figure \ref{stshweak} the critical shear rate $a^2_c$ for the 
{\em even} solution calculated from Eq.(\ref{evendet}) using
the material parameters of MBBA ($b=0.58$) is shown 
as a function of $1/w_a$, $1/w_p$.

\begin{center}
[Insert figure \ref{stshweak} here]
\end{center}

In order to obtain an {\em easy-to-use} expression for the critical shear
rate of the relevant {\em even} mode as a function of the surface
anchoring strengths one can use the single-mode approximation in the
spirit of a Galerkin expansion \cite{gal}.
We choose 
\begin{eqnarray}
\begin{array}{lcl}
V_y &=& C_1 \sin(2 \pi z) \; , \\
N_x &=& C_2 [w_a \cos(\pi z)+\pi] \; , \\
N_z &=& C_3 [w_p \cos(\pi z)+\pi] \; ,
\end{array}
\label{sh1mode}
\end{eqnarray}
which satisfy boundary conditions (\ref{bcvynxnz}).
Substituting this ansatz into (\ref{eqnvynxnz}) where ${\tilde S}=a^2$ and solving 
the algebraic system for $C_i, \{i=1,2,3\}$ obtained after projecting the first
of Eqs.(\ref{eqnvynxnz})  onto the first mode of Eqs.(\ref{sh1mode}), the
second equation onto the second mode {\em etc.}, one gets 
\begin{eqnarray}
\begin{array}{lcl}
a_c^2&=&3\pi^3 \sqrt{w_a w_p (4+w_a)(4+w_p)/(p_1 p_2)} \; , \\
p_1&=&w_a w_p+4(w_a+w_p)+2\pi^2 \; , \\
p_2&=&16 w_a w_p(1-b)+9\pi^2 b p_1 \; .
\end{array}
\label{even1mode}
\end{eqnarray}
The single-mode approximation (\ref{even1mode}) gives almost
the same values for $a^2_c$ as (\ref{evendet}) (relative error is smaller
than 1\%), so that they could not be  differentiated in figure \ref{stshweak}.

%
\section{Poiseuille flow}
\label{pf}
In the case of Poiseuille flow the shear rate ${\tilde S}$ is given by 
Eq.(\ref{baspois}) and the symmetry properties of 
system (\ref{eqnvynxnz}) give us the following two types of solutions: 
\begin{eqnarray*}
\;\mbox{I:}&& 
\{ V_y, N_x \mbox{ - odd}; N_z \mbox{ - even} \}
\mbox{ - ``odd'' solution }, \\
\mbox{II:}&&
\{ V_y, N_x \mbox{ - even}; N_z \mbox{ - odd} \}
\mbox{ - ``even'' solution }.
\end{eqnarray*}
The {\em odd}  solution corresponds to the {\em splay}-mode and
the {\em even} solution to the {\em twist}-mode in the notation of
Dubois-Violette and Manneville \cite{mdv1}.
Integration of the first equation in system (\ref{eqnvynxnz})
gives
\begin{equation}
\label{vy_z}
V_{y_{,z}} = K - (1-b) a^2 z N_x \; ,
\end{equation}
with $K=0$ for the {\em even} solution. For the {\em odd} solution
the integration constant $K$ must be nonzero (except for $b=1$, see below). 
After eliminating $V_{y_{,z}}$ from the third
equation in (\ref{eqnvynxnz}) one arrives at
\begin{eqnarray}
\begin{array}{lcl}
a^2 z N_z - N_{x_{,zz}} &=&0 \; , \\
a^2 z N_x - N_{z_{,zz}} &=&K \; .
\end{array}
\label{puassyst}
\end{eqnarray}
Following \cite{mdv1} we perform the transformation $Z=a^{2/3}z$ and
introduce new variables ${\mathcal U}=N_x+N_z$, ${\mathcal V}=N_x-N_z$, 
leading to
\begin{eqnarray}
\begin{array}{lcl}
{\mathcal U}_{,ZZ} - Z{\mathcal U} &=& -K a^{-4/3} \; , \\
{\mathcal V}_{,ZZ} + Z{\mathcal V} &=& K a^{-4/3} \; .
\end{array}
\label{uvsyst}
\end{eqnarray}
The general solution of (\ref{uvsyst}) can be expressed in terms of Airy
functions   $Ai(Z)$, $Bi(Z)$, $Gi(Z)$ \cite{Abramowitz}.

Let us first consider the case of the {\em even} solution ($K=0$). 
Then one has
\begin{eqnarray}
\begin{array}{lcl}
N_x= C_1[ Ai(Z)+Ai(-Z) ] +C_2 [ Bi(Z)+Bi(-Z) ]  \; , \\
N_z= C_1[ Ai(Z)-Ai(-Z) ] +C_2 [ Bi(Z)-Bi(-Z) ]  \; ,
\end{array}
\label{puassol}
\end{eqnarray}
where $C_1$, $C_2$ are constants to be determined from the boundary 
conditions (\ref{bcvynxnz}).
Substitution of (\ref{puassol}) into (\ref{bcvynxnz}) gives the
criterion for the threshold  shear rate
\begin{eqnarray}
\frac{w_a Ai^{+} + a_c^{2/3} Ai^{+}_{,Z}}
     {w_a Bi^{+} + a_c^{2/3}Bi^{+}_{,Z}}
=\frac{w_p Ai^{-} + a_c^{2/3} Ai^{-}_{,Z}}
      {w_p Bi^{-} + a_c^{2/3}Bi^{-}_{,Z}} \; .
\label{puasdet}
\end{eqnarray}
Here
\begin{eqnarray}
Ai^{\pm}=Ai\left(\frac12 a_c^{2/3}\right) \pm 
         Ai\left(-\frac12 a_c^{2/3}\right)\; , \quad  
Bi^{\pm}=Bi\left(\frac12 a_c^{2/3}\right) \pm 
         Bi\left(-\frac12 a_c^{2/3}\right) \;  
\end{eqnarray}
and similar for $Ai^+_{,Z}$ and $Bi^+_{,Z}$.
The limit of strong anchoring ($w_a \to \infty$, 
$w_p \to \infty$) gives the result of Dubois-Violette and Manneville 
\cite{mdv1}
\begin{eqnarray}
\frac{Ai(\frac12 a_c^{2/3})}{Bi(\frac12 a_c^{2/3})}=
\frac{Ai(-\frac12 a_c^{2/3})}{Bi(-\frac12 a_c^{2/3})} \; ,
\nonumber
\end{eqnarray}
leading to $a^2_c=102.59$.
The fact that $N_x$ and $N_z$ have
different $z$-symmetry leads in Eq.(\ref{puasdet}) to an asymmetry in the
dependence of the critical shear rate on $w_a$, $w_p$, 
in contrast to
the case of Couette flow. 
The critical shear rate [Eq.(\ref{puasdet})]
retains a finite value when the polar anchoring strength $w_p$
(which mainly controls $N_z$ perturbations) vanishes, 
whereas $a^2_c \to 0$ if the azimuthal anchoring strength $w_a \to 0$.

For the {\em odd}  solution with nonzero $K$ the solution of
(\ref{puassyst})  has the following form
\begin{eqnarray}
\begin{array}{lcl}
N_x &=& k K\{ C_1 Ai^{-}(Z) + C_2 Bi^{-}(Z) + Gi^{-}(Z)\} \; , \\
N_z &=& k K\{ C_1 Ai^{+}(Z) + C_2 Bi^{+}(Z) + Gi^{+}(Z)\} \; ,
\end{array}
\label{oddsol}
\end{eqnarray}
where 
\begin{eqnarray}
k=\pi a^{-4/3}/2, \quad Ai^{\pm}(Z)=Ai(Z) \pm Ai(-Z) \mbox{ {\em etc., } for
$Bi^{\pm}(Z)$ and $Gi^{\pm}(Z)$} \; .
\end{eqnarray}
The coefficients $C_i$ in (\ref{oddsol}) are determined from the boundary
conditions (\ref{bcvynxnz}).
Integrating Eq.~(\ref{vy_z}) and taking into 
account the boundary conditions $V_y(z=\pm 1/2)=0$ one obtains
the expression for the critical shear rate of the {\em odd}  mode
\begin{eqnarray}
K-(1-b)a^2_c
\int_{-1/2}^{1/2} z N_x(z;a^2_c,w_a,w_p) dz=0 \; .
\label{anoddsol}
\end{eqnarray}
Since $N_x$ is proportional to $K$, this undetermined integration constant 
falls out from (\ref{anoddsol}). 
From thermodynamical conditions \cite{lesadv,plbrand} the parameter $b$ has to
be positive, but is otherwise not restricted. The point $b=1$ requires special
consideration. In this case, from Eq.(\ref{vy_z}) and boundary conditions for
$V_y$, it follows that $K=0$ and $V_y=0$, so that one should solve
Eqs.(\ref{puassyst}) keeping the symmetry of $N_x$ and
$N_z$ corresponding to the {\em odd} solution. In other words, if $b=1$ then
one has the homogeneous instability of {\em odd} type with zero velocity
perturbation, like in the case of Fr\'eedericksz transition. Moreover, in the
case $b=1$ Eqs.(\ref{puassyst}) became invariant with respect to change
$\{N_x \Leftrightarrow N_z\}$, so that the critical shear rates for {\em even}
and {\em odd} solutions are the same up to transposition
$\{w_a~\Leftrightarrow~w_p\}$. 

The instability
of the {\em odd}  mode is mainly controlled by the  polar anchoring strength
$w_p$. In the limit of zero  azimuthal anchoring strength $w_a \to 0$ one has
a finite critical shear rate, whereas $a^2_c \to 0$ for the polar anchoring
strength $w_p \to 0$.

The critical shear rate $a^2_c$ for the {\em even} solution 
[Eq.(\ref{puasdet})] and {\em odd}  solution [Eq.(\ref{anoddsol})]
have been calculated for the material parameters of MBBA for various
values of $w_a$, $w_p$ (figure  \ref{stps}).
Depending on the azimuthal and polar surface anchoring strengths
one can have different $z$-symmetry of the first unstable mode.
The critical line in ($w_a,w_p$) space corresponds to the
crossover between the two types of unstable solutions.

\begin{center}
[Insert figure \ref{stps} here]
\end{center}

Since the expressions (\ref{puasdet}) and (\ref{anoddsol}) for the 
critical shear rates of the two possible unstable modes are quite
complicated we use a single-mode Galerkin approximation in order to 
obtain {\em easy-to-use} formulas.

For the {\em even} mode we can use Eqs.(\ref{puassyst}) with $K=0$. 
Assuming
\begin{eqnarray}
\begin{array}{lcl}
N_x &=& C_1 [w_a \cos(\pi z) + \pi] \; , \\
N_z &=& C_2 [w_p \sin(2\pi z) + 2\pi \sin(\pi z)] \; , 
\end{array}
\label{puasanzeven}
\end{eqnarray} 
after substituting (\ref{puasanzeven}) into (\ref{puassyst}) and projection one
obtains
\begin{eqnarray}
\begin{array}{lcl}
a^2_c&=&6\sqrt{3} \pi^4 
  \sqrt{w_a (4+w_a)(3\pi^2+20w_p+3 w_p^2)/p_1^2} \; , \\
p_1&=&16 w_a w_p+9\pi^2(w_a+w_p)+72\pi^2\; .  
\end{array} 
\label{puassceven}
\end{eqnarray}
For the {\em odd}  mode we choose 
\begin{eqnarray}
\begin{array}{lcl}
N_x &=& C_1 [w_a \sin(2\pi z) + 2\pi \sin(\pi z)] \; , \\
N_z &=& C_2 [w_p \cos(\pi z) + \pi] \; .
\end{array}
\end{eqnarray} 
Substituting this ansatz into (\ref{puassyst}) one finds the coefficients
$C_1$ and $C_2$. Then from (\ref{anoddsol}) one obtains 
\begin{eqnarray}
\begin{array}{lcl}
a^2_c&=&6\sqrt{3} \pi^4 \sqrt{w_p (4+w_p)(3\pi^2+20 w_a+3w_a^2)/(p_1 p_2)} \; ,\\ 
p_2&=&b p_1 + (1-b)(9\pi^2 -144 - 2 w_a )w_p \; .
\end{array}
\label{puasscodd}
\end{eqnarray}
The approximate expressions (\ref{puassceven}) and (\ref{puasscodd}) give
systematically higher values for the critical shear rate compared 
to (\ref{puasdet}) and (\ref{anoddsol}), about 10\%
for both modes.
Equating (\ref{puassceven}) and (\ref{puasscodd}) one can obtain
an approximate expression for the critical line in the ($w_a,w_p$) 
plane corresponding to the crossover between the critical  
{\em even}  and {\em odd}  modes (see figure \ref{stps}). 
We now write $\beta_a=1/w_a, \beta_p=1/w_p$.
For strong azimuthal anchoring ($\beta_a=0$) the transition from
{\em even}  to {\em odd}  mode takes place at some critical value of
polar anchoring strength: 
$\beta_p=\beta_{p0}$, where $\beta_{p0}$ is the solution of the
algebraic equation
\begin{eqnarray}
\begin{array}{ll}
&9\pi^4 b \beta_{p0}^3+2\pi^2 (39b-19)\beta_{p0}^2+ \\
&[-9\pi^2(1-b)+120b-\frac{232}{3}]\beta_{p0}-18(1-b)=0 \; .
\end{array}
\end{eqnarray}
For MBBA material parameters one finds $\beta_{p0}=0.307$.
Assuming the deviation  ($\hat{\beta}_a,\hat{\beta}_p$) from ($0,\beta_{p0}$)
to be small one has up to first order in
$\hat{\beta}_a$, $\hat{\beta}_p$ for the critical line
\begin{eqnarray}
c_1 (\beta_{p0}+\hat{\beta}_p) + c_2 \hat{\beta}_a = 0 \; ,
\label{alxalz}
\end{eqnarray}
where
\begin{eqnarray}
\begin{array}{lcl}
c_1&=&27 b \pi^4 \beta_{p0}^2 + 4\pi^2(39b-19)\beta_{p0}-
9\pi^2(1-b)+120b-\frac{232}{3} \; , \\ 
c_2&=&108 b \pi^4 \beta_{p0}^3 + \pi^2(9b\pi^2+936b-680)\beta_{p0}^2-
36(3\pi^2+40)(1-b)\beta_{p0}+ \\ 
&+& 216b-\frac{776}{3} \; .
\end{array}
\end{eqnarray}
The results for the critical line for the crossover between
{\em even} and {\em odd} critical modes obtained from the numerical
solutions of Eqs.(\ref{puasdet}), (\ref{anoddsol}) and 
approximation (\ref{alxalz}) are shown in figure~\ref{comp}.
\begin{center}
[Insert figure \ref{comp} here]
\end{center}
Note, that $\beta_a=1/w_a$, $\beta_p=1/w_p$ are 
inversely proportional to the
thickness $d$ of the NLC layer. 
Therefore, varying the cell thickness, one can cross the critical line
separating the two regimes.
For that purpose it is necessary to have the ratio 
$\beta_p/\beta_a  \equiv K_{11} W_a/(K_{22} W_p)$
larger than the slope of the critical line 
($\sim -c_2/c_1$).
Using the material parameters of MBBA one obtains
$W_a/W_p > - c_2 K_{22}/(c_1 K_{11})=2.28$  ( $W_a/W_p > 1.78$
from numerical results).
For small values of $w_a$ and $w_p$, the crossover is given by
$w_a=w_p/b$. As was said above, if $b=1$, Eqs.(\ref{puassyst}) are 
invariant with respect to $\{N_x \Leftrightarrow N_z\}$. This means that 
the crossover line for $b=1$ is exactly defined by $w_a=w_p$. The same result
follows from approximation formulas (\ref{puassceven}) and (\ref{puasscodd}).
%

%
\section{Conclusions and discussion}
\label{disc}
It was found that changes of the anchoring strengths can cause a crossover
between two types of homogeneous instabilities induced by steady Poiseuille 
flow, in contrast to the case of steady Couette flow, where the first
unstable mode is always the {\em even} one.
Semi-analytical expressions for the critical shear rates for both
Poiseuille and Couette flow are presented, together with
simple approximate formulas of good accuracy.

The effect of the reduction of the critical shear rate for homogeneous
instabilities under weak surface anchoring can in principle be used for the
determination of the polar ($W_p$) and azimuthal ($W_a$) anchoring 
strengths.
For that purpose one may measure the critical shear rates for two
cells with different thicknesses and use the full expressions
(or approximate formulas) to calculate both $W_p$ and $W_a$.
Alternatively, one can use an additional electric field applied across the
NLC cell.
Depending on the sign of the dielectric anisotropy of the NLC the
electric field will stabilise or destabilise $n_z$ fluctuations and affect 
the critical shear rate differently for different values 
of the polar and azimuthal anchoring strengths.
This study is in progress.
The advantage of using measurements of the critical shear rate
for the determination of polar and azimuthal anchoring strengths
compared to the other methods (Fr\'eedericksz transition, orientational 
transition with hybrid
orientation, small-angle light scattering) is that here one
does not need to modify (or rebuild) the experimental setup
in order to obtain $W_p$ and $W_a$ at the same time.

It would be particularly interesting to investigate experimentally the
orientational behaviour of NLC under steady Poiseuille flow for a
cell with $W_p$, $W_a$ close to the crossover line separating critical modes
of different symmetry.

\section*{Acknowledgements}

Financial support of 
Deutscher Akademischer Austauschdienst (DAAD),
DFG (Grant Kr690/14-1) and INTAS (Grant  96-498)   
are gratefully acknowledged. O.T. and A.K. wish to thank the 
University of Bayreuth for its hospitality.

\pagebreak


\begin{figure}[h]
\begin{center}
\parbox{13cm}{
\epsfxsize=13cm
\epsfysize=13cm
}
\end{center}
\caption{Critical shear rate of Couette flow {\em vs.} anchoring strengths.
$b=0.58$ (MBBA).} 
\label{stshweak} 
\end{figure}

\begin{figure}[h]
\begin{center}
\parbox{13cm}{
\epsfxsize=13cm
\epsfysize=13cm
}
\end{center}
\caption{Critical shear rate of Poiseuille flow {\it vs.} 
anchoring strengths. $b=0.58$ (MBBA).}
\label{stps}
\end{figure}

\begin{figure}[h]
\begin{center}
\parbox{13cm}{
\epsfxsize=13cm
\epsfysize=13cm
}
\end{center}
\caption{Critical line separating {\em odd} and {\em even} 
regimes in Poiseuille flow: numerical solution (solid) and 
approximation (\ref{alxalz}) (dotted). $b=0.58$ (MBBA).}
\label{comp}
\end{figure}


\pagebreak
\pagestyle{empty}

\begin{figure}[t]
\begin{center}
\parbox{13cm}{
\vspace*{2cm}
\epsfxsize=13cm
\epsfysize=13cm
\epsfbox{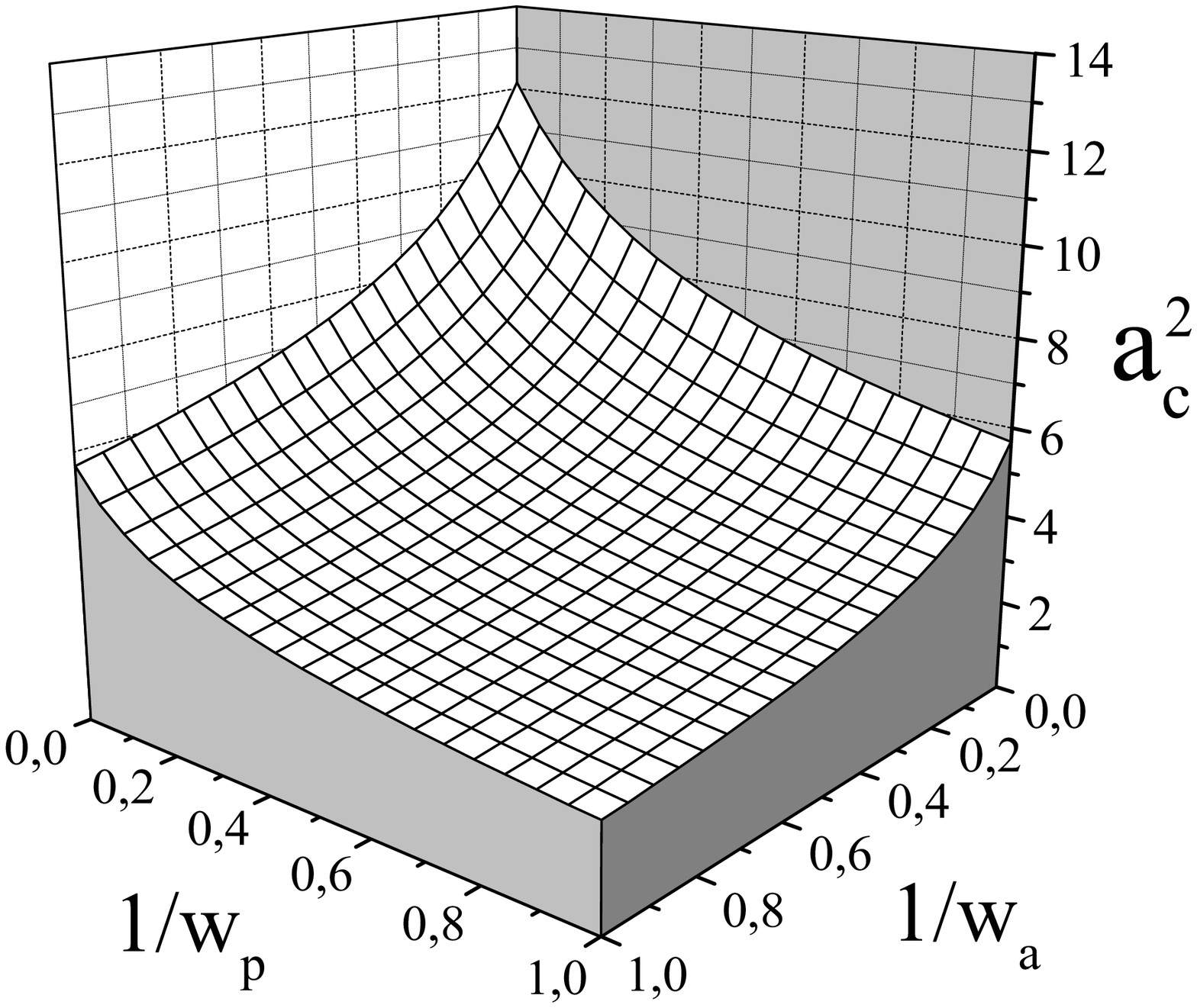}}
\end{center}
\vspace*{-0.5cm}
\end{figure}

\pagebreak

\begin{figure}[t]
\begin{center}
\parbox{13cm}{
\vspace*{2cm}
\epsfxsize=13cm
\epsfysize=13cm
\epsfbox{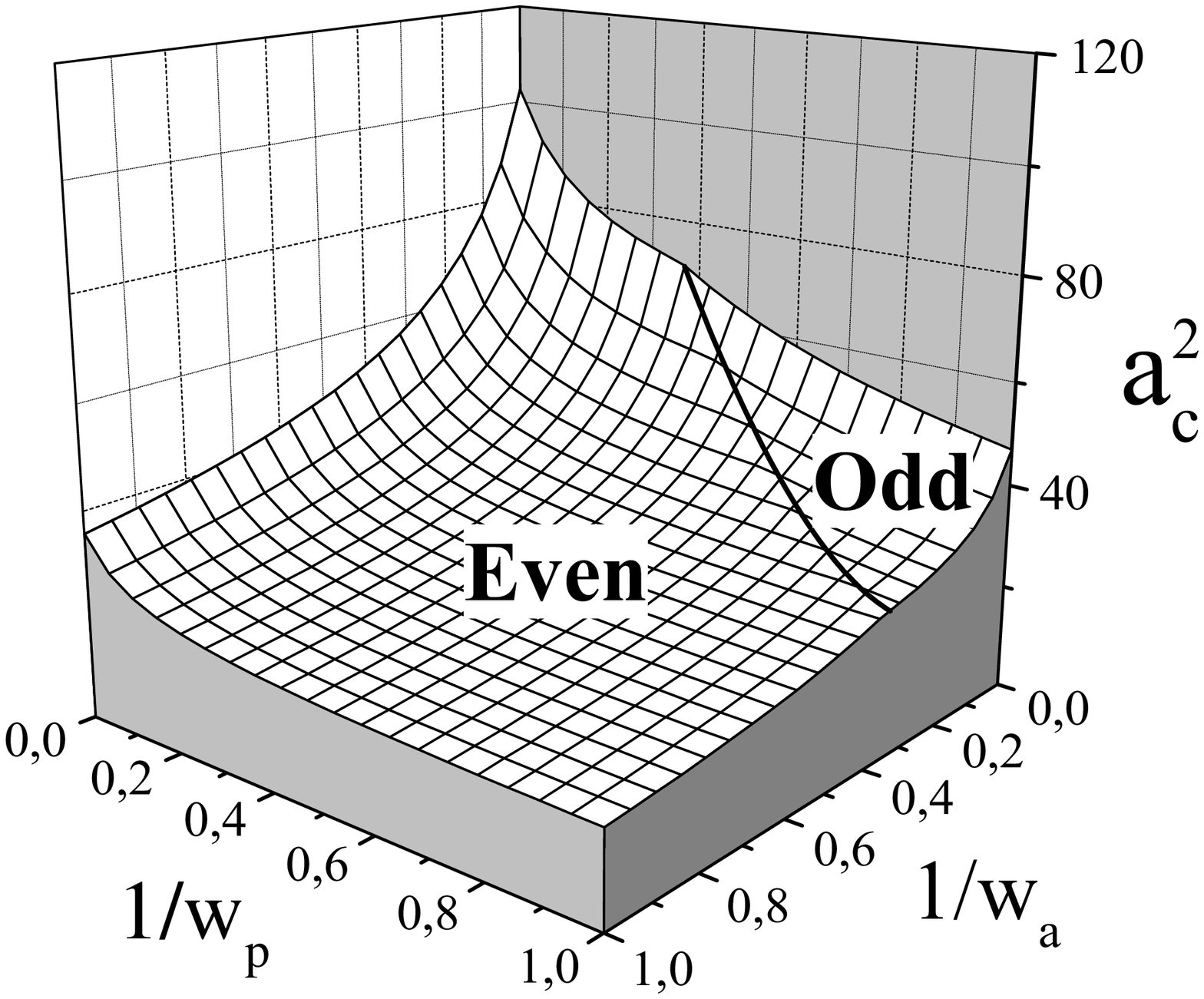}}
\end{center}
\vspace*{-0.5cm}
\end{figure}

\pagebreak

\begin{figure}[t]
\begin{center}
\parbox{13cm}{
\epsfxsize=13cm
\epsfysize=13cm
\epsfbox{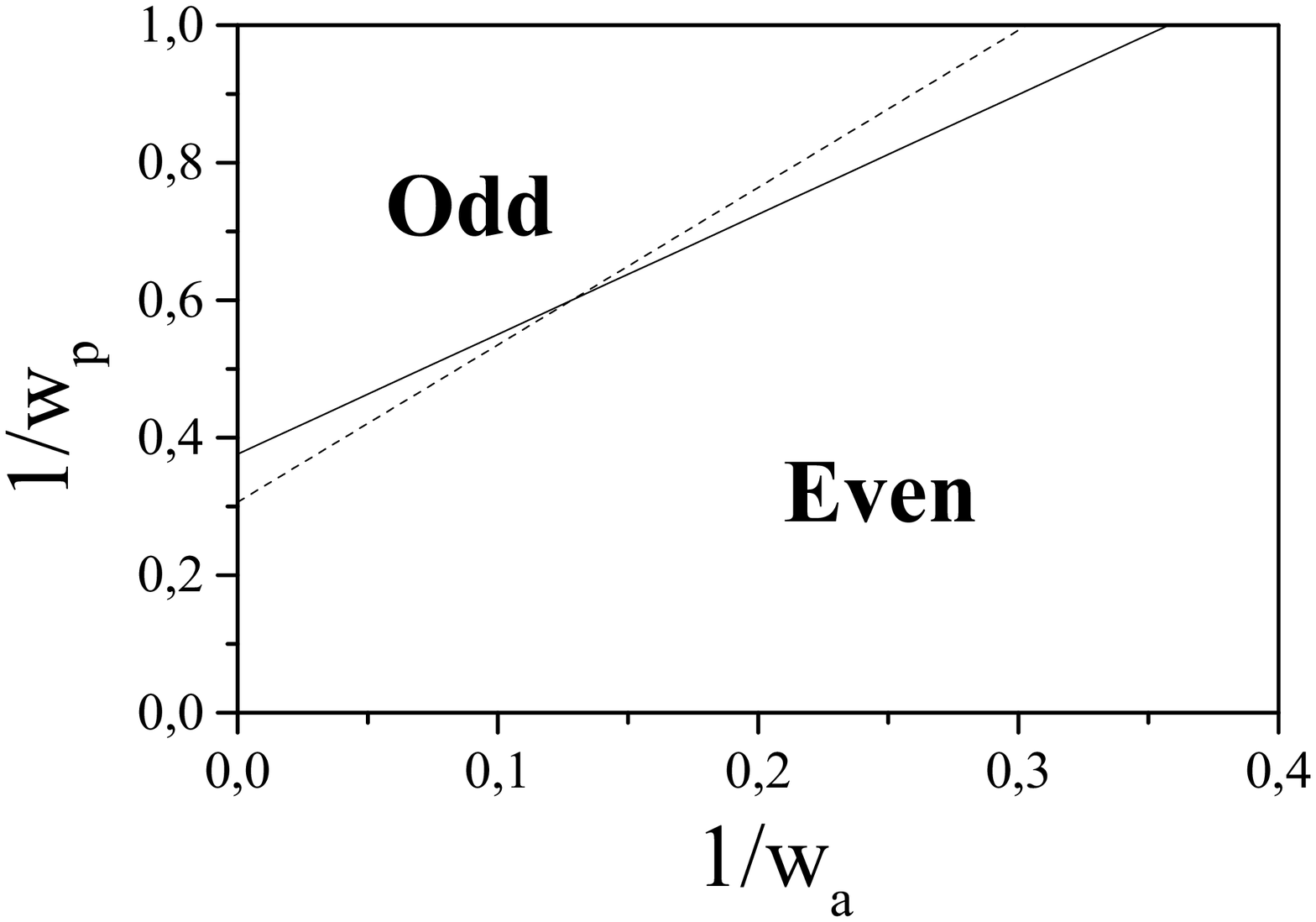}}
\end{center}
\vspace*{-0.5cm}
\end{figure}

\end{document}